\documentclass[aps,prl,twocolumn,groupedaddress]{revtex4}

\usepackage{graphicx}

\begin{document}


\title{Origin of half-semimetallicity induced at interfaces of C-BN heterostructures}


\author{J. M. Pruneda}
\email[]{miguel.pruneda@cin2.es}
\affiliation{Centre d'Investigaci\'on en Nanoci\'encia i Nanotecnolog{\'{\i}}a (CSIC-ICN). Campus de la UAB, E-08193 Bellaterra, Spain}


\date{\today}

\begin{abstract}
First-principles density functional calculations are performed in C-BN heterojunctions.
It is shown that the magnetism of the edge states in zigzag shaped graphene strips and 
polarity effects in BN strips team up to give a spin asymmetric screening that induces 
{\it half-semimetallicity} at the interface, with a gap of at least a few tenths of eV 
for one spin orientation and a tiny gap of hundredths of eV for the other. 
The dependence with ribbon widths is discussed, showing that a range of ribbon widths 
is required to obtain half-semimetallicity.
These results open new routes for tuning electronic properties at nanointerfaces and 
exploring new physical effects similar to those observed at oxide interfaces, in lower 
dimensions.
\end{abstract}


\maketitle

\section{}
A number of new physical phenomena have been discovered in the last few years at the 
interfaces between very different materials. The electronic reconstruction induced at 
these boundaries can give rise to metallic states\cite{Hwang}, magnetism\cite{Brinkman}, 
or even superconductivity\cite{Mannhart}, although the parent compounds were originally 
insulating oxides.  But not only boundaries of bulk materials are important: in two 
dimensional graphene nanoribbons (GNR) edges have become relevant, with peculiar 
electronic states localized at the boundary of the ribbons\cite{Fujita}.

Since its first experimental realization\cite{Novoselov}, graphene has emerged as 
a prominent candidate to replace silicon in the development of high performance 
electronic devices.  However, the extremely high mobility of charge carriers in 
graphene (ten times higher than that in silicon wafers used in microprocessors) 
poses a difficulty for the fabrication of nanodevices based on this material.  
A possible route to harness the conducting charges in graphene transistors is 
the fabrication of graphene nanoribbons, where the lateral confinement of charge 
induces the opening of a gap that depends on the orientation of the GNR, and is 
inversely proportional to its width\cite{Han,Son-gap}. Interestingly, ribbons 
with zigzag shaped border (ZGNR) present a local ferrimagnetic structure at the edge\cite{Fujita}, 
and upon application of an in-plane external electric field can be tuned into a 
half-metallic state, with a gap opening for one spin component, and a metallic 
behavior for the other, hence giving a full spin polarization of the conducting 
electrons\cite{Son-halfmetal}. Theoretical calculations have shown that similar 
magnetic edge states exist in zigzag edged BN nanoribbons (ZBNNR)\cite{Barone, Zheng}. 
The recent experimental realization of BN nanosheets\cite{Pacile,Iijima}, opens new 
prospects for the combination of these two isostructural materials. Hence, a proper 
characterization of these hypothetical semimetal-insulator junctions is much needed.

Different C/BN heterostructures have been proposed in the past, and substantial efforts 
have been devoted to the growth of composite sheets and nanotubes\cite{reviewCBN}. 
Laser vaporization grown C-BN single walled nanotubes show traces of the patterning of 
segregated BN nanodomains embedded in the carbon network sequentially along the tube 
axis\cite{Enouz}. Recently, Du {\it et. al.} used first principles Molecular Dynamics 
calculations to prove that hybrid C-BN nanotubes can be spontaneously formed via the 
connection of BNNR and GNR at room temperature\cite{Du}, and Ding {\it et. al.} 
investigated the stability of C doped BNNR, showing that, under suitable conditions, 
GNR can be grown embedded in BN sheets\cite{Ding}. Furthermore, it is reported that 
half-metallicity originates for sufficiently wide ZGNRs\cite{Ding}. Tuning 
half-metallicity by edge modification in ZGNR and ZBNNR has been demonstrated by 
others\cite{Zheng,Kan,Dutta,Wu}, and understood in terms of a potential difference 
between the two edges through chemical modification with electron accepting or donating 
groups.  If this was the explanation for the half-metallicity in GNR embedded in BNNR, 
then there should be half-metallicity independently of the BNNR width. This doesn't seem 
to be the case, as the results in reference \onlinecite{Dutta} suggest that there is a critical 
thickness for the BNNR to induce half-metallicity in GNRs. 


Here, I argue that the electronic properties of C+BN heterostructures can be 
described in terms of the 1D equivalence of bidimensional oxide heterojunctions made 
of a metal (semimetal graphene) and an insulator (BN sheet). In particular, it will be 
shown that the polarity discontinuity in BN nanoribbon and the screening by the mobile 
electrons in graphene are behind the electronic reconstruction at the interface. Hence, a
dependence with the BNNR width (polarity) and GNR width (number of available screening 
carriers) is expected.

Ab initio pseudopotential density functional calculations are performed in zigzag 
edged Graphene-BN superlattices, within the spin-polarized generalized-gradient 
approximation as implemented in the SIESTA method\cite{siesta}. Troullier-Martin 
type pseudopotentials\cite{pseudos} and numerical atomic orbitals with double-$\zeta$ 
plus polarization are used to describe the electronic valence states. The atomic 
positions are determined with a structural relaxation until the forces are smaller 
than 0.02eV/Å. A total of 67 k-points are used to sample the Brillouin zone. 
Following the conventional nomenclature $n$-ZGNR and $m$-ZBNNR combine to give a 
$(n,m)$-superlattice with $n$ zigzag chains of graphene and $m$ zigzag chains of BN (Fig. 1a).  
A set of systems with $2\le n\le12$ and $2\le m\le11$ are considered, with 
superlattice's periodicities in the range 1.7-4.7nm. The smaller systems ($n+m=8$) 
show non-magnetic (NM) semiconducting properties, with a direct energy gap that 
decreases with the increase in the width of the graphene ribbon.  For wider systems however the 
gap is reduced and the system semimetallic, with edge states that give rise to a narrow band 
near the Fermi level, enabling the possibility of magnetic orderings induced by 
the electron-electron interaction\cite{Fujita}. Indeed, spin-polarized edge 
states are obtained in the calculations, that correspond to antiferromagnetic 
(AF) alignments of the spin moments at opposite edges of the heterostructure 
(Fig. \ref{bands}), with the AF state being a few meV 
lower than NM, in agreement with calculations of magnetic edge states in isolated 
ZGNR, and ZGNR embedded in ZBNNR\cite{Ding}. 
Double periodicity unit cells were used to explore other magnetic orderings observed
for isolated BNNR\cite{Barone},
but the AF-phase remained as the ground state. Notice that hybrid exchange-DFT 
functionals predict a further stabilization of the AF alignment in isolated ZGNR
\cite{Pisani}. Notice also the half-semimetallic character of the AF state in 
Fig. 2, with an apparent gap for $\alpha$ spin, and two bands that seem to cross 
the Fermi level for $\beta$ spin. The crossing is not allowed by symmetry because 
each sublattice of C atoms is linked either to B or N, breaking the sublattice's 
equivalency and giving a tiny gap also for the $\beta$-spin bands. Hence, the term
{\it half-semimetal} is more appropriate than {\it half-metal} commonly used in 
the literature.

 \begin{figure}
\includegraphics[width=0.45\textwidth]{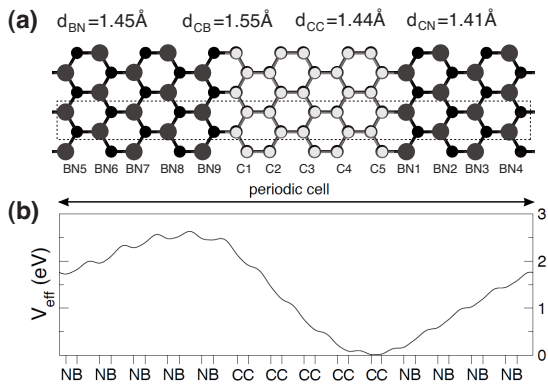}
 \caption{\label{structure}(a) Diagram of a (5,9) C-BN superlattice with 5 and 9 zigzag 
chains for the C-ribbon and BN-ribbon, respectively. White, black, and dark gray circles 
represent C, N and B atoms. The dashed line box shows the cell of repetition under periodic 
boundary conditions. Calculated C-C, B-N, C-B and C-N bond lengths are also shown. 
(b) Profile of the macroscopic electronic effective potential averaged over the graphene 
plane. The origin of the potential was shifted to the N-edge of the superlattice.}
 \end{figure}

 \begin{figure}[b]
 \includegraphics[width=0.45\textwidth]{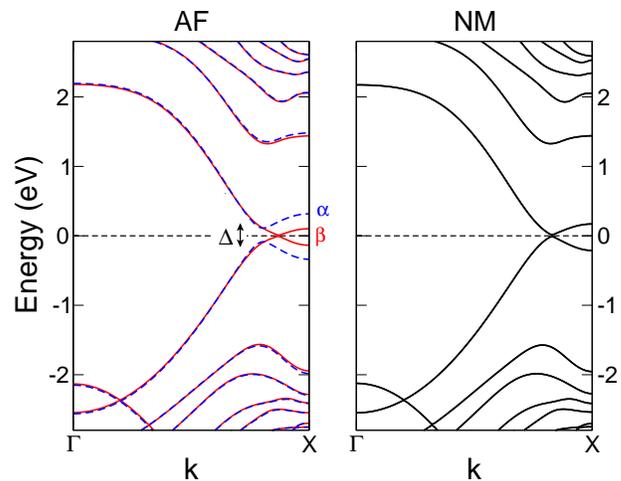}
 \caption{\label{bands} (color online) Electronic bandstructure of C-BN superlattices 
in the antiferromagnetic (AF) groundstate and non-magnetic (NM) state for the 
(5,9)-superlattice. Dashed (blue) and solid (red) lines correspond to $\alpha$ and 
$\beta$ spins. The presence of an energy gap, $\Delta$, for the $\alpha$ spin is 
apparent in the AF state. }
 \end{figure}

The nature of these edge bands near the Brillouin zone boundary (X-point at k=$\pi$) 
deserves further clarification. Hydrogen passivated zigzag GNRs have insulating magnetic 
edge states with parallel ordering at each edge, and antiparallel alignment between the 
two edges.  On the other hand, zigzag BNNRs have insulating, non-magnetic edge states 
with unoccupied state localized at the B site and occupied valence state localized at 
the N side. These states can be dramatically affected by edge passivation\cite{Lee,Zheng}. 
When GNR and BNNR are attached the mixing of $\pi$ orbitals of C, B and N at each 
edge gives rise to four sets of bands that correspond to the bonding and antibonding 
states between C-N and C-B (Fig. 3a). The relevant bands, close to the Fermi level, 
are the occupied bonding C-B ($\pi_B$), and unoccupied antibonding C-N ($\pi_{N}^*$), 
each of them being a localized state at carbon atoms close to the B or N edge, 
respectively. Figure 3.b-e presents evidences of the highly localized states at 
X by showing a color contour plots for the energy bands near the Fermi level projected 
on the C atomic orbitals at the atoms close to the N and B edges. The localization 
gradually decreases as we depart from the X-point. The bonding C-N and antibonding C-B 
lay $\sim$5eV below and above the Fermi level respectively. Ab initio calculations 
have shown that the bandgaps in both C and BN zigzag nanoribbons are inversely 
proportional to their width, and consequently the energy separation between $\pi_B$ 
and $\pi_{N}^*$ in C-BN superlattices will depend on the width of each strip. 
Hence, a dependence of the half-semimetallic gap on the widths of the strips is 
anticipated, and will be discussed later.  

 \begin{figure}
 \includegraphics[width=0.45\textwidth]{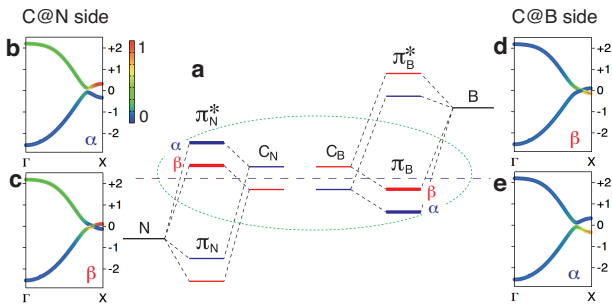}
 \caption{\label{Elevels}(color online) (a) Energy diagram for localized states close 
to the C-N (left) and C-B (right) edges, for $\alpha$ (blue) and $\beta$ (red) spins. 
GNR edge states have opposite spin character at each side of the ribbon, and couple 
to the B and N orbitals. Near the Fermi level (dashed horizontal line), the relevant 
states are occupied $\pi_B$ and unoccupied $\pi_{N}^*$, with some spin asymmetry coming 
from the original GNR spin edge polarization.  The weights on the relevant bands of 
the C atomic orbitals at the N and B edges are shown in b, c, d and e for each spin, 
with red/blue being a high/negligible contribution to the electronic wavefunction. 
The Fermi level is taken at 0 and the bands are plotted along the $\Gamma$-X direction.}
 \end{figure}

The different electronegativities of B and N make bare ZBNNR the two-dimensional 
equivalent to a polar slab, with a type 3 termination, where alternating charges are 
ordered perpendicularly to the edge, as opposed to armchair BNNR which have type 1 
boundaries and alternating charges parallel to the interface\cite{Tasker}. 
Polarization lowers the potential felt by electrons at the nitrogen edge relative 
to the boron edge.  Charge redistributions near the interface partially compensate 
this edge instability, particularly for wider ribbons, where the number of electrons 
that can participate in the screening is higher\cite{Park-Louie}. 
Atomic relaxations also contribute to the screening, with bonds becoming shorter 
at the N-side and longer at the B-side.  It is then expected 
that mobile electrons coming from semimetallic graphene in contact with these ZBNNR 
will increase the screening. But this also means that there is a charge asymmetry in 
the GNR in contact with BN with the interface C-N being lower in energy than the C-B 
interface, and hence inducing a charge transfer to the N side. This charge 
transfer is confirmed by Mulliken population analysis, with a slightly larger 
charge for C atoms close to the N interface, than for C atoms 
close to the B edge. This result is at odds with the hypothesis that the origin of the
half-metallicity in the GNR is the charge transfer from the carbon to boron atoms\cite{Dutta}.
Although the occupied states close to the Fermi level are mainly localized at the atoms 
close to this side (Fig. 3), the fact that for very narrow ZGNR half-semimetallicity 
is not observed\cite{Ding} proves that this postulate is incorrect.

The situation very much resembles that of a GNR under the presence of a transverse 
electric field, except that the nature of the edge states is different, with occupied bands 
near the Fermi level localized at the same edge (B-side) for both spin polarizations. 
Figure 1b shows the macroscopic average\cite{Baldereschi} of the electronic effective 
potential in the plane of the (5,9)-superlattice. This potential profile shows a drop 
of $\sim$2.3eV between B and N edges, inducing an effective electric field of 
$\sim$0.27V/\AA, which is at the brink of the critical electric field needed to induce 
half-metallicity in the isolated ZGNR\cite{Son-halfmetal}. The potential drop increases 
slightly with the BN widths studied here but the decrease of the effective polarization 
charge density ensures that it will saturate for sheets wide enough. For wider graphene 
strips, the screening is asymmetric and mainly localized a few (4 or 5) zigzag chains 
from the B-edge, so that the effective electric field inside the graphene ribbon will 
be substantially screened far from this edge.  This means that the widths of 
the GNR and BNNR strips can be used to tune the electronic and magnetic properties of 
these systems.  

The most important result of this work is 
shown in Fig. 4 where the band gaps for $\alpha$ and $\beta$ spins are plotted as a 
function of the strip widths. In the limit of broad GNR, the semimetallic behavior is 
recovered, with zero gap for both spins.  If the GNR is too thin, the AF state is unstable 
and the systems becomes NM and insulating.  However, lattices with sufficiently wide BN strips 
($m\ge6$) and sufficiently narrow C ribbons ($n\le8$) can show half-semimetallic 
properties with a gap for $\alpha$-spin states significantly larger (above 0.1eV) 
than for $\beta$-spin (below 0.03eV). It is important to notice that the origin of this effect
is not simply the chemical modification of the ZGNR by B or N doping, as the spin polarization
disappears for narrow BN strips (inset in Fig. 4).

 \begin{figure}
 \includegraphics[width=0.45\textwidth]{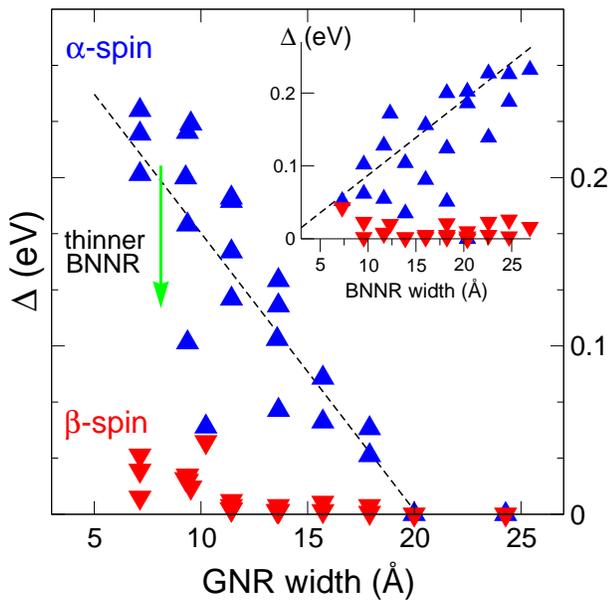}
 \caption{\label{gaps} (color online) Direct band gaps for $\alpha$- (blue up triangles) 
and $\beta$-spins (red down triangles) as a function of the GNR width in the 
antiferromagnetic ground state for C+BN superlattices. The (green)  arrow shows the 
dependence of the $\alpha$-spin gap on the BNNR width (also displayed in the inset). 
Superlattices made from GNR thinner than 15Å and BNNR thicker than 10Å present a marked 
half-semimetallicity with a ≥0.1eV gap for $\alpha$-spins and a tiny gap for $\beta$-spins.}
 \end{figure}

As mentioned before, hybrid C-BN nanotubes have already been synthesized experimentally\cite{Enouz}, 
and predictions have been made for the spontaneous formation of single-walled armchair 
nanotubes from the hybrid connection of BNNR and GNR\cite{Du}.  Unzipping these hybrid 
nanotubes would be a route to fabricate C-BN superlattices as the ones reported in this work. 
However, it can be shown that even in the tubular geometry the electronic 
and magnetic properties described above can be obtained, as long as there is a zigzag 
edge and the widths of BN and C ribbons are conveniently tuned, following the 
gap width-dependence sketched in Fig. 4. Curvature does not play an essential role, and 
$(n,n)$ zigzag nanotubes with $n\ge 5$ can show these half-metallic properties. Details will 
be presented elsewhere\cite{EPAPS}.

In addition to the demonstration that unconventional interfacial effects are not limited 
only to planar junctions in epitaxial complex oxides heterostructures but are present in 
lower dimensions, synthesis of these C-BN nanostructure offers an interesting avenue for 
the design of carbon-based nanospintronic devices. A zigzag-edged nanostriction of C-BN in 
a graphene ribbon, for example, would open a gap and induce a high spin-polarization for 
the transmitted electrons without the need for applied external electric fields, paving 
the way to efficient spin-injection in carbon structures. 
Surely different combination of materials can be conceived to fabricate bidimensional 
heterostructures and take advantage of these unusual physical phenomena that result from 
the combination of spin-polarized edge states and polarization induced charge dipoles at 
the edges. Further work on growing bidimensional superlattices that combine the exiting 
possibilities of nanosheets is encouraged. 

\begin{acknowledgments}
This work was supported by the CSIC through a JAE-Doc fellowship and P.I.E. 200960I025. 
Computational resources from the Theory and Simulation group are acknowledged.  
The author is thankful to P. Ordej\'on, E. Artacho and E. Canadell for fruitful 
discussions and comments.
\end{acknowledgments}

\bibliography{biblio}
\end{document}